%% file: Bohnet_HybridSuperradSensor_2012.tex
\newcommand{\ket}[1]{\ensuremath{\left|  #1 \right\rangle}}
\newcommand{\expec}[1]{\ensuremath{\left\langle {#1} \right\rangle}}

\documentclass[prl,aps,reprint,showpacs]{revtex4-1}
\usepackage{graphicx}
\usepackage{bibunits}

\begin{document}
\begin{bibunit}

\title{Dynamically controlled superradiant laser for hybrid sensing of
collective atomic coherence}
\author{Justin G. Bohnet}
\email[Send correspondence to: ]{bohnet@jilau1.colorado.edu}
\author{Zilong Chen}
\author{Joshua M. Weiner}
\author{Kevin C. Cox}
\author{James K. Thompson}
\affiliation{JILA, NIST and Department of Physics, University of Colorado,\\ Boulder, Colorado 80309-0440, USA}


\begin{abstract}

We implement dynamic control of a superradiant, cold atom $^{87}$Rb Raman laser to realize the equivalent of conditional Ramsey spectroscopy for sensing atomic phase shifts.  Our method uses the non-demolition mapping of the collective quantum phase of an ensemble of two-level atoms onto the phase of a detected cavity light field. We show that the fundamental precision of the non-demolition measurement can theoretically approach the standard quantum limit on phase estimation for a coherent spin state, the traditional benchmark for Ramsey spectroscopy. Finally, we propose a hybrid optical lattice clock based on this method that combines continuous and discrete measurements to realize both high precision and accuracy.

\end{abstract}

\maketitle

Dynamically tunable interactions between atoms are essential to a wide range of recent advances including quantum logic gates\cite{Wineland95, Saffman10}, the realization of ultracold polar molecules\cite{Ni2008}, and quantum simulations of quantum many-body systems\cite{Blatt2012, Bloch2012}.  In the field of precision measurement, interactions can lead to collective effects that improve precision.  Examples in which phase resolution is enhanced through interactions include interferometry using soliton matter waves\cite{KSF02,SPG02} and spin-squeezed states of matter (including ions\cite{Wineland01}, thermal gases\cite{AWO09,SLV10,LSM10,CBS11}, and quantum degenerate gases\cite{GHM08, GZN10}). As another example, short range collisional interactions between atoms can create spin locking for enhanced coherence times for improving precision \cite{DRL10}.  Similarly, the spontaneous synchronization of atomic dipoles, mediated by the cavity field in masers\cite{SCT58,GKR60} and superradiant lasers\cite{MYC09, BCW12} generate collective coherence times greater than single particle coherence times.

\begin{figure}[t]
\includegraphics[width=3.5in]{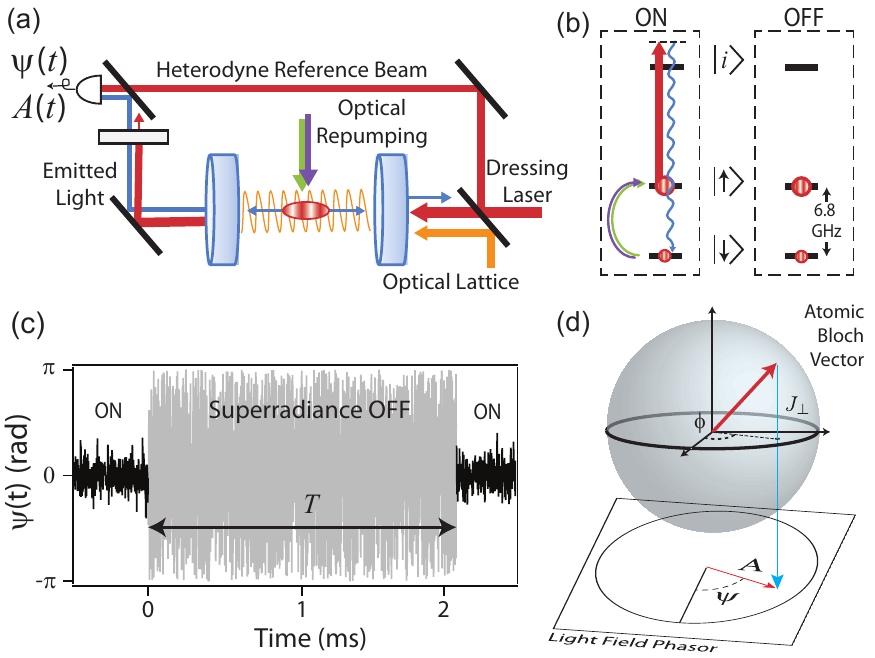}
\caption{(color online) (a) The experimental setup and (b) a simplified atomic energy level diagram. $^{87}$Rb atoms (red) are trapped in a 1D optical lattice (orange) within an optical cavity.  A Raman dressing laser (red) detuned from an intermediate state $\ket{i}$ induces an optical decay at rate $\gamma$ between ground hyperfine states $\ket{\uparrow}$ and $\ket{\downarrow}$. The emitted light (blue) is resonant with a single cavity mode.  The atoms are incoherently repumped back to $\ket{\uparrow}$ at a rate $w$ using two repumping lasers (green and purple).  When the dressing and repumping lasers are off, the atoms remain in a superposition of \ket{\uparrow} and \ket{\downarrow}, with a quantum phase $\phi(t)$ that evolves at $6.834$ GHz. (c) Emitted light phase $\psi(t)$. The superradiant emission continuously maps $\phi(t)$ onto $\psi(t)$. We measure a differential light phase $\psi(T)-\psi(0)$ during a free evolution time $T$ using dynamic control of $\gamma$ and $w$. (d) Graphic representation of mapping of the collective atomic Bloch vector\cite{FVH57} $\vec{J}$onto the phasor representing the emitted light field.}
\label{ExptSetup}
\end{figure}

However, interactions can also cause systematic errors and long term instabilities, e.g., collisional shifts in atomic clocks\cite{Saloman02} or matter wave interferometers \cite{PCC01}. Dynamically tunable interactions offer the benefit over static interactions in that they can be turned on to first create desired collective phenomena and then later turned off to allow high accuracy measurements, free from perturbations associated with both the interactions and the control mechanism.  Here, we provide a proof-of-principle  demonstration and analysis of how cavity mediated interactions can be controlled to create long coherence times, enable high precision non-demolition measurements of atomic coherence, and enable high accuracy, Ramsey-like measurements in a single system. 

Specifically, this Letter consists of three new results.  First, we provide a proof-of-principle experimental demonstration of a novel atomic sensor based on a dynamically tunable optical Raman superradiant laser  between ground states of  $^{87}$Rb (Fig. \ref{ExptSetup}a and \ref{ExptSetup}b) \cite{BCW12}.  Because the atoms act as the primary reservoir of phase information in this laser, we show that the active oscillation can be interspersed at-will with periods of passive atomic phase evolution to sense an applied phase shift (Fig. \ref{ExptSetup}c).  The mode of operation (passive or active) is determined solely by whether cavity-mediated optical interactions between the atoms are enabled by an externally applied Raman dressing laser. The hybrid active/passive sensor operates quasi-continuously with a high repetition rate, because the atomic phase accumulated during a passive period is mapped onto the phase of the light emitted in subsequent periods of active oscillation (Fig. \ref{ExptSetup}d) in a manner that does not destroy the ensemble, as is often the case for fluorescence detection\cite{Biedermann09}.

Secondly, we provide a theoretical analysis of the fundamental limitations on phase estimation sensitivity for an ideal three-level superradiant atomic sensor. The analysis shows that the output light provides sufficient information to continuously track the evolving phase of the atomic coherence with a precision near the standard quantum limit (SQL) on phase resolution for a coherent spin state $(\Delta \phi_{SQL})^2 = 1/N$.  This result is conceptually important for fundamental understanding of how quantum noise sources (specifically Schawlow-Townes phase diffusion and vacuum noise) combine to limit our fundamental knowledge of the instantaneous value of the collective atomic phase. In addition, the result demonstrates that little fundamental precision needs to be sacrificed for the future hybrid oscillators proposed here, relative to utilizing traditional Ramsey spectroscopy\cite{PhysRev.78.695}.
 
Thirdly, we propose how the model system we demonstrate could be realized in an optical lattice clock \cite{Katori2011} with the goal of building a clock with both high short term stability derived from high measurement bandwidth during active oscillation and high accuracy due to periods of passive oscillation with low systematic perturbations.  Such an approach may be important for moving high precision optical clocks out of carefully controlled laboratory environments for diverse applications \cite{LTB11}.

This paper is related to other work on engineering and dynamic control of the coupling between large ensembles of atoms\cite{BDR07, TGS09, MR09, TWL11} and single atoms\cite{Dayan08, SNR11} in an optical cavity. This work is also related to the development of non-demolition\cite{AWO09,SLV10,CBS11} and sample-preserving\cite{LWL09} measurement techniques for neutral atoms that may improve measurement precision and  mitigate aliasing noise\cite{Westergaard2010} in atomic clocks.  The storage of phase information in the atomic ensemble presented here is closely related to the field of quantum memories for quantum information\cite{DLCZ01,BBM07,SNR11,CGP10, RDZ10,LMR11}. 

\begin{figure}[t]
\includegraphics[width=3.5in]{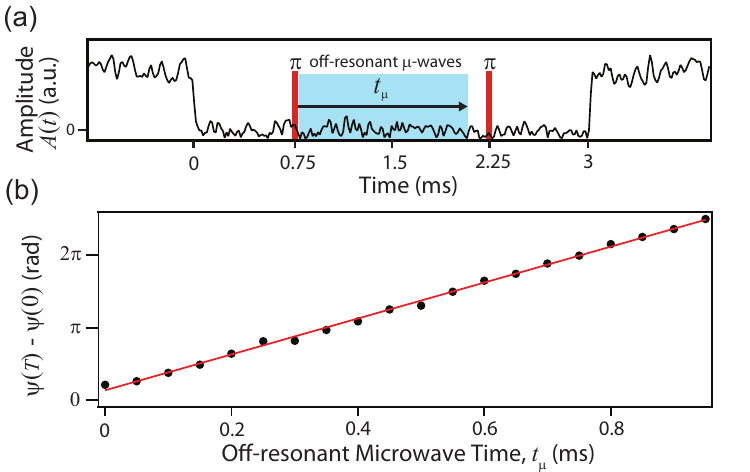}
\caption{(color online) (a) The emitted light amplitude as a function of time showing a spin-echo sequence with superradiant phase readout used to sense an applied phase shift. The phase shift is induced by microwaves detuned from the $\ket{\downarrow}$ to $\ket{\uparrow}$ transition near 6.834~GHz. (b) The measured phase shift grows linearly with the time for which the microwaves are applied $t_\mu$, and the fitted slope is in reasonable agreement with the predicted value (red line is a fit).}
\label{Phase Measurement}
\end{figure}

We perform our proof-of-principle experiments using a cold atom, $^{87}$Rb Raman laser (see Fig. \ref{ExptSetup}, Ref. \cite{BCW12}). The laser operates deep into the bad-cavity, or superradiant, regime\cite{PhysRevLett.72.3815,PhysRevA.47.1431} where the cavity power decay rate $\kappa/2 \pi  = 11$~MHz is much larger than all other relevant decay and scattering rates in the system.   The laser utilizes $N=10^6$ $^{87}$Rb atoms trapped in a 1-D optical lattice at $823$~nm inside a moderate finesse $F\approx700$ optical cavity that also serves to mediate the long-range interactions between the atoms that drive spontaneous synchronization of the atomic dipoles.

Dynamic control of the interactions is achieved through a tunable optical Raman dressing laser, operating at 795~nm, inducing a decay rate $\gamma/2\pi \approx 1$ Hz from $\ket{\uparrow} = \ket{F=2, m_F=0}$ to $\ket{\downarrow} =\ket{F=1, m_F=0}$ (Fig. \ref{ExptSetup}b). The dressing laser is tuned 1~GHz to the blue of an optically excited intermediate state $\ket{i} = \ket{F'=2}$, and the rate at which a single atom would scatter photons from this laser into the resonant cavity mode is $\Gamma_c = C\gamma$, where $C = 8 \times 10^{-3}$ is the single-particle cavity cooperativity parameter of cavity-QED\cite{BCW12}.  The collective cooperativity satisfies the necessary condition for superradiance $NC \gg 1$. Single-particle repumping is accomplished at an optimum repumping rate\cite{MYC09} $w \approx w_{pk} = N\Gamma_c/2$ using separate lasers at 780~nm. All lasers can be switched on and off within 100~ns -- much faster than the timescale on which atomic dynamics occur. 

The ensemble of atoms can be represented by a Bloch vector $\vec{J}$ (Fig. \ref{ExptSetup}d) whose azimuthal phase $\phi(t)$ evolves in time at a rate set by $\dot{\phi}(t) = E(t)/\hbar$, where $E(t)$ is the instantaneous energy difference separating the two levels and $\hbar$ is the reduced Planck constant.  Depending on the sensitivity of $E(t)$ to environmental conditions, precise measurements of $\phi(t)$ correspond to measurements of the environment or time\cite{Kitching2011}.

The quantum phase $\phi(t)$ is not directly measurable and must be mapped onto an observable quantity.  Bad-cavity active oscillators, such as masers or the present superradiant laser, continuously map the collective phase $\phi(t)$ onto the observable phase $\psi(t)$ of an electromagnetic field (Fig. \ref{ExptSetup}d), because the rapidly decaying cavity field is slaved to the atomic coherence.  Ignoring vacuum noise, the complex electric field phasor is given by $A(t) e^{i \psi(t)}\propto J_\perp(t) e^{i \phi(t)}$ where $J_\perp(t)$ is the projection of the Bloch vector onto the x-y plane and $A(t)$ is the amplitude of the electric field phasor.  The optical phase measurement is equivalent to a continuous non-demolition measurement of the evolving atomic coherence, and can provide much higher measurement bandwidth over passive Ramsey-like evolution as occurs in atomic fountain clocks or optical lattice clocks.

To realize the equivalent of a passive Ramsey free evolution period, we perform the measurement sequence in Fig. \ref{Phase Measurement}. We apply the dressing and repumping lasers to start steady-state superradiance for some period of time.  We measure the phase $\psi(t)$ and amplitude $A(t)$ of the emitted light via heterodyne detection with respect to the dressing laser to remove the phase noise on the dressing laser.  By splitting the resulting RF signal and simultaneously demodulating the two quadratures, we obtain both $A(t)$ and $\psi(t)$ simultaneously.  

After the system has settled to steady state, we estimate the light phase $\bar{\psi}(0)$  just before we temporarily shut off the dressing and repumping lasers at a time we define as $t=0$. We make the estimate using linear fits to 0.5~ms of $\psi(t)$ data at times $t<0$.  After $t=0$, the superradiant emission turns off, but the collective Bloch vector continues to precess at the microwave frequency $6.834$ GHz.  At time $t=T$, the repumping and dressing lasers are turned back on, and the phase of the light is once again estimated as $\bar{\psi}(T)$ using $\psi(t)$ data at times $t>T$.  The evolved atomic phase during the time period $T$ is then estimated as  $\phi(T)-\phi(0) = \bar{\psi}(T) -\bar{\psi}(0)$.

To demonstrate the sensing of an applied phase shift, we applied off-resonance microwaves that shift the $\ket{\uparrow}$ to $\ket{\downarrow}$ transition frequency for variable amounts of time $t_{\mu}$ during the passive evolution period. The time required to create large phase shifts necessitated the removal of background dephasing from the optical lattice trap, so we applied two spin-echo $\pi$-pulses using a separate microwave source.  

The differential light phase $\bar{\psi}(T) - \bar{\psi}(0)$ versus $t_{\mu}$ is shown in Fig. \ref{Phase Measurement}b.  The slope is in reasonable agreement with the theoretical prediction for the atomic frequency shift caused by the applied microwaves. Unlike in traditional Ramsey spectroscopy, there is no sinusoidal fringe because we measure phase, not atomic population. Thus, there is no need to operate near an optimal bias point because the light's phase can be measured with essentially equal sensitivity regardless of its value. 

\begin{figure}[t]
\includegraphics[width=3.5in]{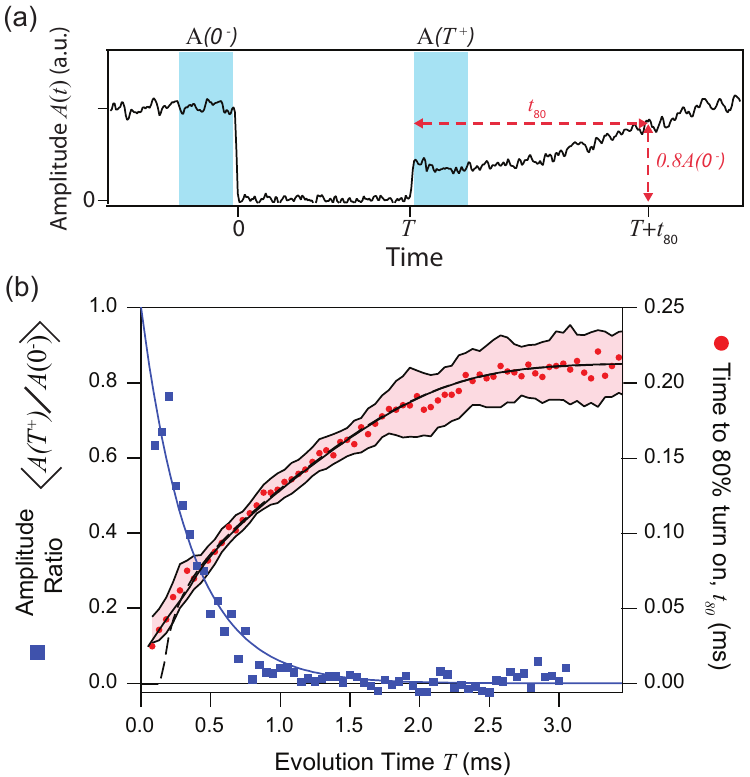}
\caption{(color online) (a) An average time trace of the emitted light amplitude returning to steady state after the loss of coherence during the evolution time $T$. $A(0^-)$ and $A(T^+)$ are average amplitudes measured in the blue time windows before and after the evolution time.  The return to steady state is quantified with a characteristic rise time $t_{80}$ at which the amplitude has returned to 80\% of its steady state value. (b, left) The decay of the atomic coherence as estimated by the ratio $A(T^+)/A(0^-)$ (blue squares).  The blue curve is the loss of coherence $c(t)$ as determined from an independent fit to the decay of contrast using traditional Ramsey microwave spectroscopy.  
(b, right) Measured (red circles) and predicted (dashed line) recovery time $t_{80}$. The theoretical prediction uses the measured contrast decay function $c(t)$ as an input and calculates $t_{80}$ with a semi-classical rate equation model. The solid black curve is the prediction accounting for the low-pass filter that was applied to the data.  The band around the data indicates 1 s.~d. on each side of the data point. We attribute the fluctuations observed at short times to finite measurement precision.  Each point is the average of 20 trials.}
\label{TurnOn}
\end{figure}

Next, we observe how the system behaves versus loss of coherence. We can observe the decay of the atomic coherence in the dynamics of re-establishment of superradiance at $t=T$. In Fig. \ref{TurnOn}, we measure the amplitude of the emitted light field just before turn off $A(0^-)$ at $t=0$ and just after turn on $A(T^+)$ at $t=T$. When the evolution time $T$ is short, the superradiance promptly returns to $A(0^-)$, but as the atomic coherence $J_{\perp}$ decays during the evolution time, $A(T^+)$ decreases proportionally. The ratio of amplitudes $A(T^+)/A(0^-)$ shown in Fig. \ref{TurnOn}b is well described by the fitted fringe contrast $c(T)$ as measured using traditional microwave Ramsey spectroscopy with population readout.

The lost coherence during $T$ is eventually restored as the oscillator returns to steady state.  In Fig. \ref{TurnOn}b, the time $t_{80}$ at which $A(T+t_{80})/A(0^{-})= 0.8$ is measured and compared to a numerical theoretical prediction based on the observed atomic contrast $c(T)$.  At short times, little coherence is lost, and the oscillator quickly reaches steady state after turn on because the remaining coherence provides a seed for the superradiance.  

At very long times, the remaining atomic coherence is small. With no seed to restart the oscillation, the system must wait for a photon to be spontaneously scattered into the cavity mode to re-establish the coherence. The random nature of spontaneous emission also results in larger fluctuations in $t_{80}$ at large $T$ (Fig. \ref{TurnOn}b). We observe fluctuations of the same order as the predicted fluctuations in the time to reach the peak intensity of a superradiant pulse after preparation in the fully excited state\cite{Gross1982}.

While the coherence times in this proof-of-principle experiment are limited by technical imperfections, such as differential lattice shifts, our experiment confirms the role of the collective atomic coherence in the hybrid sequence and demonstrates that the re-seeded light has a well defined emission phase, even in the presence of finite loss of coherence.

\begin{figure}[t]
\includegraphics[width=3.5in]{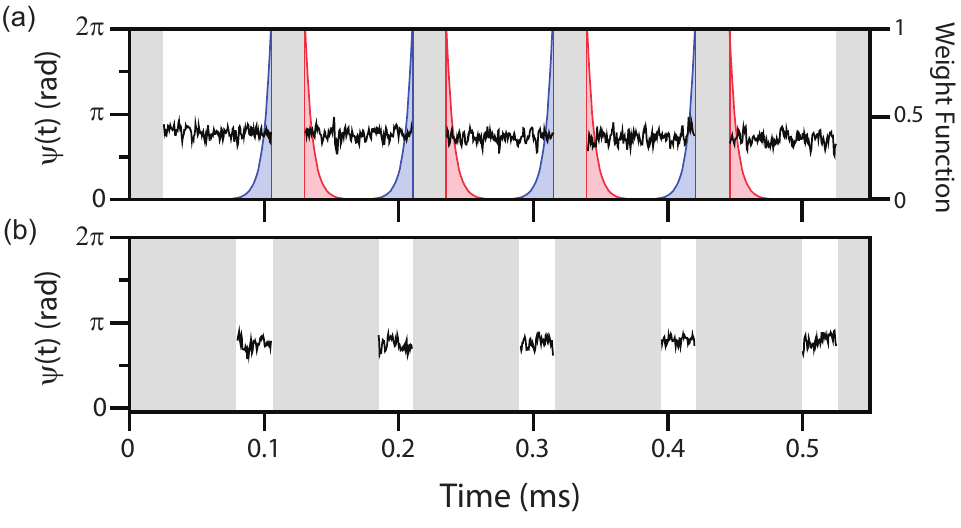}
\caption{(color online). Demonstration of repeated dynamic control of superradiance for hybrid active/passive oscillator.  The measured light phase $\psi(t)$ is shown when the superradiance is switched on (black).  The periods of evolution time, when the superradiance is switched off are shown as gray regions (random phase data not shown).  The duty cycle can be chosen at will to minimize systematic errors at long time scales versus improving the measurement bandwidth. Parts (a) and (b) are two examples.  Ideally, optimal estimates $\phi_e(t)$ of the desired underlying atomic phase $\phi(t)$ are obtained using an exponential weight function with time constant $\tau_W$ as shown by the blue and red curves shown in the first graph. The blue and red curves correspond to the weight functions before and after an evolution period respectively.  The weighting function here is calculated for an ideal superradiant light source but with our experimental parameters of $N$, $\Gamma_c$, and $q$.}
\label{Blink}
\end{figure}

We also demonstrate coherence-preserving measurements by repeating on/off sequences to create a hybrid active/passive phase reference, shown in Fig. \ref{Blink}.  The free evolution times would ideally have high accuracy, while the active oscillation would have much greater measurement bandwidth. In Fig. \ref{Blink} (a) and (b) we show two example experimental trials where only the duty cycle has changed.  While the details are beyond the scope of the present work, the duty cycle of the measurements could be adjusted in real time for optimal overall phase stability and accuracy given knowledge of the environment. This might allow such a robust oscillator to be employed outside of the laboratory for both scientific and commercial applications.

Having demonstrated the measurement technique in a model system, we now consider the fundamental quality of the continuous non-demolition mapping of atomic phase $\phi(t)$ onto the field phase $\psi(t)$ both in terms of the precision of the mapping and the rate at which information is gained (i.e. measurement bandwidth). 

The quality of the mapping is limited by fundamental quantum noise in the form of Schawlow-Townes phase diffusion of the atomic coherence and photon shot noise on the measurement of the light phase.  Employing a Kalman filter\cite{Kalman1961} analysis we find that the optimal estimate $\phi_{e}(t)$ is an exponentially weighted average of $\psi(t)$ with a weighting time constant $\tau_W = 1/(\sqrt{q}N\Gamma_c)$, assuming $w=w_{pk}$ and where $q$ is the photon detection efficiency\cite{SOM}. A single-pole, low pass filter could be used to implement such a running weighted average (Fig. \ref{Blink}a).

The mean squared error of the optimal estimator is $\sigma_{e}^2 = \left< \left(\phi_e\left(t\right) - \phi\left(t\right)\right)^2 \right> = \frac{2}{\sqrt{q} N}$.  When $q=1$, the error is within a factor of 2 of the standard quantum limit (SQL) on phase variance for a coherent spin state of $N$ unentangled atoms $(\Delta \phi_{SQL})^2$, a natural benchmark connecting to traditional Ramsey spectroscopy\cite{SOM}.  The impact of finite photon detection efficiency $q<1$ is mitigated compared to what one would expect for the increase in the measurement photon shot noise alone.  This is because photon losses can be partially compensated by an increased weighting constant $\tau_W$, a parallel result to that of quantum non-demolition measurements recently demonstrated in several systems\cite{AWO09,SLV10,CBS11}.  

The Raman system presented in this work does not realize the predicted fundamental precision, primarily due to dispersive tuning of the cavity mode as described in Ref. \cite{BCW12}.  However, the dispersive tuning of the cavity mode is not a flaw inherent to the tunable superradiant oscillator scheme, so these results should not be interpreted as a limitation to a future active/passive hybrid sensor.

Finally, we propose that future hybrid oscillators can be implemented on ultranarrow optical transitions.   A number of schemes for inducing tunable decay rates $\gamma$ of the excited state of alkaline-earth elements with strictly forbidden transitions have been demonstrated or proposed\cite{BHO06,SAI05}.  As an example, a static magnetic field can induce a tunable decay rate $\gamma$ on the strictly forbidden transition $^3P_0$ to $^1S_0$ in $^{174}$Yb, as experimentally demonstrated in a passive optical lattice clock\cite{BHO06}. A decay linewidth of $\gamma/2\pi=0.2$~Hz can be induced by a $0.1~$Gauss magnetic field that could be switched on and off quite rapidly $\ll 1$~ms.  Assuming a cavity cooperativity $C = 0.1$, $N=10^5$, and $q=1$, the weighting time $\tau_W$ is only $80~\mu$s, faster than both the obtainable single-particle decoherence time scales that limits $T$ in a Ramsey-like experiment and the time needed to measure the populations using fluorescence detection.  Alternatively, schemes that dynamically change the finesse of the cavity, such as $Q$ switching, may also provide a dynamic control mechanism.

We have realized a proof-of-principle hybrid atomic sensor in a superradiant Raman laser, theoretically analyzed the fundamental quality of the non-demolition mapping of atomic coherence to light coherence, and proposed an implementation of a hybrid sensor in an optical lattice clock. Though the system demonstrated in this work is of limited use for precision measurement, it nevertheless points a way forward to developing a new class of atomic sensors and highlights unique properties for precision measurement that emerge from many-body systems with dynamically tunable interactions.

We thank Jun Ye, Murray Holland, Dominic Meiser, and David Tieri for enlightening conversations.  All authors acknowledge financial support from NSF PFC, NIST, and ARO.  J.G.B. acknowledges support from NSF GRF, and Z.C. acknowledges support from A*STAR Singapore.

\end{bibunit}

\begin{bibunit}
\include{Bohnet_HybridSuperradSensorSOM_2012}

\end{bibunit}

\end{document}

%% file: Bohnet_HybridSuperradSensorSOM_2012.tex
\title{Dynamically controlled superradiant laser for hybrid sensing of
collective atomic coherence: supporting online material}

\author{Justin G. Bohnet}
\email[Send correspondence to: ]{bohnet@jilau1.colorado.edu}
\author{Zilong Chen}
\author{Joshua M. Weiner}
\author{Kevin C. Cox}
\author{James K. Thompson}
\affiliation{JILA, NIST and Department of Physics, University of Colorado,\\ Boulder, Colorado 80309-0440, USA}


\maketitle

\subsection{Derivation of Optimal Estimator}
Here we derive the optimal estimator $\phi_e(t)$ of the quantum phase $\phi(t)$, and its mean squared error $\sigma^2_e$, given a measurement record $\psi(t)$ of the phase of the superradiantly emitted optical field.  We use the results from a continuous Kalman filter analysis with uncorrelated process noise and measurement noise\cite{Kalman1960, Kalman1961, zarchan_fundamentals_2001}. Here the measurement noise corresponds to the photon shot noise that appears in the measurement of the light phase $\psi(t)$ and the process noise corresponds to the phase diffusion of the collective atomic Bloch vector that sets the Schawlow-Townes frequency linewidth limit. 

\vspace{3mm}
{\bf Photon Shot Noise}
The measured phase of the radiated light is related to the underlying quantum phase $\phi(t)$ by, $\psi(t) = \phi(t) + \Delta\psi(t)$ where the vacuum or photon shot noise adds the noise component $\Delta\psi(t)$. The noise is Poissonian and described by its lowest order moments as $\left <  \Delta\psi(t) \right> =0$ and  $\left <  \Delta\psi(t)\,\Delta\psi(t+\tau) \right> =\delta(\tau)\Phi_m$.  Here, $\Phi_m$ is the power spectral noise density of phase fluctuations, and $\delta(\tau)$ is the Dirac delta function so that the measurement noise at different times is uncorrelated.  The light phase variance for a measurement time $\Delta t$ is $\sigma_m^2 = \Phi_m/\Delta t = \frac{1}{4 R_d \Delta t}$, where $R_d$ is the average rate of detected superradiantly-emitted photons using homodyne detection.  At the optimum superradiant photon emission rate\cite{MYC09,BCW12} $R_d =  N^2 \Gamma_c /8$, obtained at a repumping rate $w_{pk} = N \Gamma_c/2$ where $\Gamma_c = C\gamma$ is the single-particle natural decay rate into the cavity mode. Taking into account finite quantum efficiency $q$, we find 

\begin{equation}
\Phi_m = \frac{2}{q N^2 \, \Gamma_c}\,.
\end{equation}

{\bf Phase Diffusion}
In addition to measurement noise, the collective Bloch vector's quantum phase $\phi(t)$ diffuses with time as a result of quantum noise in the repumping process, the same mechanism that sets the Schawlow-Townes frequency linewidth limit in a bad-cavity laser or maser\cite{PhysRevA.47.1431,PhysRevLett.72.3815,MYC09}.   As a result, values of $\phi(t)$ at different times are less correlated with one another as the time separation grows.  Specifically, the two-time phase difference (as measured in an appropriate rotating frame) averaged over many trials is zero $\expec{ \phi(t+\tau)-\phi(t) } =0$, but the variance of the phase difference grows linearly with the time difference $\tau$ as

\begin{equation}\label{DiffNoise}
\sigma^2_D(\tau)=\expec{(\phi(t+\tau)-\phi(t))^2}=D^2 \left | \tau \right |
\end{equation}

\noindent The phase diffusion coefficient $D$ for the superradiant source can be derived from the expectation value of the two-time raising and lowering atomic operator $\expec{\sigma_+(t+\tau)\sigma_-(t)}$ in Ref. \cite{MEH10} and is

\begin{equation}
D^2= \Gamma_c \left (1 + \frac{2 w}{N \Gamma_c} \right )
\end{equation}

\noindent  Assuming operation at the repumping rate $w_{pk}$ one finds $D^2 = 2 \Gamma_c$. For the Kalman filter analysis to follow, we need the spectrum of frequency fluctuations defined as 

\begin{equation}
\Phi_D = \int_{-\infty}^{\infty}\expec{\dot{\phi}(t+\tau)\dot{\phi}(t)}\cos(\omega \tau)~d\tau
\end{equation} 

\noindent which can be shown to equal a constant $D^2$.

{\bf Optimal Phase Estimation with Kalman filter}
The Kalman filter\cite{Kalman1960,Kalman1961,zarchan_fundamentals_2001} is designed to provide an optimal estimate $\phi_e(t)$ of the phase $\phi(t)$ that minimizes the mean squared error in the estimate $\sigma_e^2 = \expec{\left(\phi_e(t) -\phi(t)\right)^2}$.  The Kalman filter assumes the state model and noise sources are well known, and the that the process noise and measurement noise are uncorrelated, an assumption we verify by extending the theoretical work of Ref. \cite{PhysRevA.47.1431} to the spectrum of phase fluctuations in a homodyne measurement.  

For this simple case, the optimal Kalman filter takes the form of a single pole low-pass filter.  In the time domain, this is equivalent to an exponential weighting of the measurement record characterized by the exponential time constant $\tau_W$.  The time constant is the inverse of the Kalman gain $K$, which is calculated in steady state by a ratio of the noise spectral densities

\begin{equation}
\tau_W = \frac{1}{K} = \left( \frac{\Phi_m}{\Phi_D} \right)^{1/2} = \frac{1}{\sqrt{q}N\Gamma_c} \,,
\end{equation}

\noindent assuming $w=w_{pk}$. The optimal estimate is then the exponentially weighted average $\phi_{e} (t) = \frac{1}{\tau_W}\int_{-\infty}^{t} \psi\left ( t' \right)  e^{-(t-t')/\tau_W} \mathrm{d} t'$. 

\noindent The mean squared error in the optimal estimate is given by the geometric mean of the noise spectral densities

\begin{equation}
\sigma_e^2 = (\Phi_D \Phi_m)^{1/2} = \frac{2}{\sqrt{q}N} 
\end{equation}

Here we simply considered portions of the measurement record $\psi(t)$ at times $t \le t_\circ$, as this is the only information actually available were the superradiance to be shut off at time $t_\circ$ as part of a Ramsey-like measurement.  Conversely, an estimator of the phase just after superradiance is turned back on $\phi_e(t_\circ + T)$ will only include the measurement record at times $t\ge t_\circ+T$.  The symmetry of the two noise processes with respect to time reversal makes it sufficient to consider only the first case.

\subsection{Definition of the standard quantum limit}
The standard quantum limit (SQL) on phase measurement in the absence of any entanglement between atoms is set by quantum projection of each atom in the ensemble into either $\ket{\uparrow}$ or $\ket{\downarrow}$. In ideal Ramsey spectroscopy, the Bloch vector's projection during the dark time $T$ is maximum $J_{\perp}= N/2$ and the SQL phase measurement variance for an ensemble of $N$ atoms is $\sigma^2_{SQL}=1/N$.  After accounting for the fact that at the optimum repumping rate $w_{pk}$, the projection is reduced to $J_{\perp}=\frac{N}{2\sqrt{2}}$, the corresponding SQL for measuring $\phi(t)$ would be $\tilde{\sigma}^2_{SQL} = 2 \sigma^2_{SQL}$. In an ideal Ramsey measurement, we have perfect state preparation, but the superradiant sensor requires an initialization measurement that also has some uncertainty. The variance on the differential estimator of the superradiant measurements $\phi_e(T)-\phi_e(0)$ is then $2\sigma^2_e$.

%% file: Bohnet_HybridSuperradSensor_2012.bbl
\begin{thebibliography}{47}%
\makeatletter
\providecommand \@ifxundefined [1]{%
 \@ifx{#1\undefined}
}%
\providecommand \@ifnum [1]{%
 \ifnum #1\expandafter \@firstoftwo
 \else \expandafter \@secondoftwo
 \fi
}%
\providecommand \@ifx [1]{%
 \ifx #1\expandafter \@firstoftwo
 \else \expandafter \@secondoftwo
 \fi
}%
\providecommand \natexlab [1]{#1}%
\providecommand \enquote  [1]{``#1''}%
\providecommand \bibnamefont  [1]{#1}%
\providecommand \bibfnamefont [1]{#1}%
\providecommand \citenamefont [1]{#1}%
\providecommand \href@noop [0]{\@secondoftwo}%
\providecommand \href [0]{\begingroup \@sanitize@url \@href}%
\providecommand \@href[1]{\@@startlink{#1}\@@href}%
\providecommand \@@href[1]{\endgroup#1\@@endlink}%
\providecommand \@sanitize@url [0]{\catcode `\\12\catcode `\$12\catcode
  `\&12\catcode `\#12\catcode `\^12\catcode `\_12\catcode `\%12\relax}%
\providecommand \@@startlink[1]{}%
\providecommand \@@endlink[0]{}%
\providecommand \url  [0]{\begingroup\@sanitize@url \@url }%
\providecommand \@url [1]{\endgroup\@href {#1}{\urlprefix }}%
\providecommand \urlprefix  [0]{URL }%
\providecommand \Eprint [0]{\href }%
\providecommand \doibase [0]{http://dx.doi.org/}%
\providecommand \selectlanguage [0]{\@gobble}%
\providecommand \bibinfo  [0]{\@secondoftwo}%
\providecommand \bibfield  [0]{\@secondoftwo}%
\providecommand \translation [1]{[#1]}%
\providecommand \BibitemOpen [0]{}%
\providecommand \bibitemStop [0]{}%
\providecommand \bibitemNoStop [0]{.\EOS\space}%
\providecommand \EOS [0]{\spacefactor3000\relax}%
\providecommand \BibitemShut  [1]{\csname bibitem#1\endcsname}%
\let\auto@bib@innerbib\@empty
\bibitem [{\citenamefont {Monroe}\ \emph {et~al.}(1995)\citenamefont {Monroe},
  \citenamefont {Meekhof}, \citenamefont {King}, \citenamefont {Itano},\ and\
  \citenamefont {Wineland}}]{Wineland95}%
  \BibitemOpen
  \bibfield  {author} {\bibinfo {author} {\bibfnamefont {C.}~\bibnamefont
  {Monroe}}, \bibinfo {author} {\bibfnamefont {D.~M.}\ \bibnamefont {Meekhof}},
  \bibinfo {author} {\bibfnamefont {B.~E.}\ \bibnamefont {King}}, \bibinfo
  {author} {\bibfnamefont {W.~M.}\ \bibnamefont {Itano}}, \ and\ \bibinfo
  {author} {\bibfnamefont {D.~J.}\ \bibnamefont {Wineland}},\ }\href {\doibase
  10.1103/PhysRevLett.75.4714} {\bibfield  {journal} {\bibinfo  {journal}
  {Phys. Rev. Lett.}\ }\textbf {\bibinfo {volume} {75}},\ \bibinfo {pages}
  {4714} (\bibinfo {year} {1995})}\BibitemShut {NoStop}%
\bibitem [{\citenamefont {Isenhower}\ \emph {et~al.}(2010)\citenamefont
  {Isenhower}, \citenamefont {Urban}, \citenamefont {Zhang}, \citenamefont
  {Gill}, \citenamefont {Henage}, \citenamefont {Johnson}, \citenamefont
  {Walker},\ and\ \citenamefont {Saffman}}]{Saffman10}%
  \BibitemOpen
  \bibfield  {author} {\bibinfo {author} {\bibfnamefont {L.}~\bibnamefont
  {Isenhower}}, \bibinfo {author} {\bibfnamefont {E.}~\bibnamefont {Urban}},
  \bibinfo {author} {\bibfnamefont {X.~L.}\ \bibnamefont {Zhang}}, \bibinfo
  {author} {\bibfnamefont {A.~T.}\ \bibnamefont {Gill}}, \bibinfo {author}
  {\bibfnamefont {T.}~\bibnamefont {Henage}}, \bibinfo {author} {\bibfnamefont
  {T.~A.}\ \bibnamefont {Johnson}}, \bibinfo {author} {\bibfnamefont {T.~G.}\
  \bibnamefont {Walker}}, \ and\ \bibinfo {author} {\bibfnamefont
  {M.}~\bibnamefont {Saffman}},\ }\href {\doibase
  10.1103/PhysRevLett.104.010503} {\bibfield  {journal} {\bibinfo  {journal}
  {Phys. Rev. Lett.}\ }\textbf {\bibinfo {volume} {104}},\ \bibinfo {pages}
  {010503} (\bibinfo {year} {2010})}\BibitemShut {NoStop}%
\bibitem [{\citenamefont {Ni}\ \emph {et~al.}(2008)\citenamefont {Ni},
  \citenamefont {Ospelkaus}, \citenamefont {de~Miranda}, \citenamefont {Pe'er},
  \citenamefont {Neyenhuis}, \citenamefont {Zirbel}, \citenamefont
  {Kotochigova}, \citenamefont {Julienne}, \citenamefont {Jin},\ and\
  \citenamefont {Ye}}]{Ni2008}%
  \BibitemOpen
  \bibfield  {author} {\bibinfo {author} {\bibfnamefont {K.-K.}\ \bibnamefont
  {Ni}}, \bibinfo {author} {\bibfnamefont {S.}~\bibnamefont {Ospelkaus}},
  \bibinfo {author} {\bibfnamefont {M.~H.~G.}\ \bibnamefont {de~Miranda}},
  \bibinfo {author} {\bibfnamefont {A.}~\bibnamefont {Pe'er}}, \bibinfo
  {author} {\bibfnamefont {B.}~\bibnamefont {Neyenhuis}}, \bibinfo {author}
  {\bibfnamefont {J.~J.}\ \bibnamefont {Zirbel}}, \bibinfo {author}
  {\bibfnamefont {S.}~\bibnamefont {Kotochigova}}, \bibinfo {author}
  {\bibfnamefont {P.~S.}\ \bibnamefont {Julienne}}, \bibinfo {author}
  {\bibfnamefont {D.~S.}\ \bibnamefont {Jin}}, \ and\ \bibinfo {author}
  {\bibfnamefont {J.}~\bibnamefont {Ye}},\ }\href {\doibase
  10.1126/science.1163861} {\bibfield  {journal} {\bibinfo  {journal}
  {Science}\ }\textbf {\bibinfo {volume} {322}},\ \bibinfo {pages} {231}
  (\bibinfo {year} {2008})}\BibitemShut {NoStop}%
\bibitem [{\citenamefont {Blatt}\ and\ \citenamefont {Roos}(2012)}]{Blatt2012}%
  \BibitemOpen
  \bibfield  {author} {\bibinfo {author} {\bibfnamefont {R.}~\bibnamefont
  {Blatt}}\ and\ \bibinfo {author} {\bibfnamefont {C.~F.}\ \bibnamefont
  {Roos}},\ }\href {\doibase 10.1038/nphys2252} {\bibfield  {journal} {\bibinfo
   {journal} {Nature Physics}\ }\textbf {\bibinfo {volume} {8}},\ \bibinfo
  {pages} {277} (\bibinfo {year} {2012})}\BibitemShut {NoStop}%
\bibitem [{\citenamefont {Bloch}\ \emph {et~al.}(2012)\citenamefont {Bloch},
  \citenamefont {Dalibard},\ and\ \citenamefont {Nascimb\`{e}ne}}]{Bloch2012}%
  \BibitemOpen
  \bibfield  {author} {\bibinfo {author} {\bibfnamefont {I.}~\bibnamefont
  {Bloch}}, \bibinfo {author} {\bibfnamefont {J.}~\bibnamefont {Dalibard}}, \
  and\ \bibinfo {author} {\bibfnamefont {S.}~\bibnamefont {Nascimb\`{e}ne}},\
  }\href {\doibase 10.1038/nphys2259} {\bibfield  {journal} {\bibinfo
  {journal} {Nature Physics}\ }\textbf {\bibinfo {volume} {8}},\ \bibinfo
  {pages} {267} (\bibinfo {year} {2012})}\BibitemShut {NoStop}%
\bibitem [{\citenamefont {Khaykovich}\ \emph {et~al.}(2002)\citenamefont
  {Khaykovich}, \citenamefont {Schreck}, \citenamefont {Ferrari}, \citenamefont
  {Bourdel}, \citenamefont {Cubizolles}, \citenamefont {Carr}, \citenamefont
  {Castin},\ and\ \citenamefont {Salomon}}]{KSF02}%
  \BibitemOpen
  \bibfield  {author} {\bibinfo {author} {\bibfnamefont {L.}~\bibnamefont
  {Khaykovich}}, \bibinfo {author} {\bibfnamefont {F.}~\bibnamefont {Schreck}},
  \bibinfo {author} {\bibfnamefont {G.}~\bibnamefont {Ferrari}}, \bibinfo
  {author} {\bibfnamefont {T.}~\bibnamefont {Bourdel}}, \bibinfo {author}
  {\bibfnamefont {J.}~\bibnamefont {Cubizolles}}, \bibinfo {author}
  {\bibfnamefont {L.~D.}\ \bibnamefont {Carr}}, \bibinfo {author}
  {\bibfnamefont {Y.}~\bibnamefont {Castin}}, \ and\ \bibinfo {author}
  {\bibfnamefont {C.}~\bibnamefont {Salomon}},\ }\href {\doibase
  10.1126/science.1071021} {\bibfield  {journal} {\bibinfo  {journal}
  {Science}\ }\textbf {\bibinfo {volume} {296}},\ \bibinfo {pages} {1290}
  (\bibinfo {year} {2002})}\BibitemShut {NoStop}%
\bibitem [{\citenamefont {Strecker}\ \emph {et~al.}(2002)\citenamefont
  {Strecker}, \citenamefont {Partridge}, \citenamefont {Truscott},\ and\
  \citenamefont {Hulet}}]{SPG02}%
  \BibitemOpen
  \bibfield  {author} {\bibinfo {author} {\bibfnamefont {K.~E.}\ \bibnamefont
  {Strecker}}, \bibinfo {author} {\bibfnamefont {G.~B.}\ \bibnamefont
  {Partridge}}, \bibinfo {author} {\bibfnamefont {A.~G.}\ \bibnamefont
  {Truscott}}, \ and\ \bibinfo {author} {\bibfnamefont {R.~G.}\ \bibnamefont
  {Hulet}},\ }\href {\doibase 10.1038/nature747} {\bibfield  {journal}
  {\bibinfo  {journal} {Nature}\ }\textbf {\bibinfo {volume} {417}},\ \bibinfo
  {pages} {150} (\bibinfo {year} {2002})}\BibitemShut {NoStop}%
\bibitem [{\citenamefont {Meyer}\ \emph {et~al.}(2001)\citenamefont {Meyer},
  \citenamefont {Rowe}, \citenamefont {Kielpinski}, \citenamefont {Sackett},
  \citenamefont {Itano}, \citenamefont {Monroe},\ and\ \citenamefont
  {Wineland}}]{Wineland01}%
  \BibitemOpen
  \bibfield  {author} {\bibinfo {author} {\bibfnamefont {V.}~\bibnamefont
  {Meyer}}, \bibinfo {author} {\bibfnamefont {M.~A.}\ \bibnamefont {Rowe}},
  \bibinfo {author} {\bibfnamefont {D.}~\bibnamefont {Kielpinski}}, \bibinfo
  {author} {\bibfnamefont {C.~A.}\ \bibnamefont {Sackett}}, \bibinfo {author}
  {\bibfnamefont {W.~M.}\ \bibnamefont {Itano}}, \bibinfo {author}
  {\bibfnamefont {C.}~\bibnamefont {Monroe}}, \ and\ \bibinfo {author}
  {\bibfnamefont {D.~J.}\ \bibnamefont {Wineland}},\ }\href {\doibase
  10.1103/PhysRevLett.86.5870} {\bibfield  {journal} {\bibinfo  {journal}
  {Phys. Rev. Lett.}\ }\textbf {\bibinfo {volume} {86}},\ \bibinfo {pages}
  {5870} (\bibinfo {year} {2001})}\BibitemShut {NoStop}%
\bibitem [{\citenamefont {Appel}\ \emph {et~al.}(2009)\citenamefont {Appel}
  \emph {et~al.}}]{AWO09}%
  \BibitemOpen
  \bibfield  {author} {\bibinfo {author} {\bibfnamefont {J.}~\bibnamefont
  {Appel}} \emph {et~al.},\ }\href {\doibase 10.1073/pnas.0901550106}
  {\bibfield  {journal} {\bibinfo  {journal} {Proc. Natl. Acad. Sci.}\ }\textbf
  {\bibinfo {volume} {106}},\ \bibinfo {pages} {10960} (\bibinfo {year}
  {2009})}\BibitemShut {NoStop}%
\bibitem [{\citenamefont {Schleier-Smith}\ \emph {et~al.}(2010)\citenamefont
  {Schleier-Smith}, \citenamefont {Leroux},\ and\ \citenamefont
  {Vuleti\ifmmode~\acute{c}\else \'{c}\fi{}}}]{SLV10}%
  \BibitemOpen
  \bibfield  {author} {\bibinfo {author} {\bibfnamefont {M.~H.}\ \bibnamefont
  {Schleier-Smith}}, \bibinfo {author} {\bibfnamefont {I.~D.}\ \bibnamefont
  {Leroux}}, \ and\ \bibinfo {author} {\bibfnamefont {V.}~\bibnamefont
  {Vuleti\ifmmode~\acute{c}\else \'{c}\fi{}}},\ }\href {\doibase
  10.1103/PhysRevLett.104.073604} {\bibfield  {journal} {\bibinfo  {journal}
  {Phys. Rev. Lett.}\ }\textbf {\bibinfo {volume} {104}},\ \bibinfo {pages}
  {073604} (\bibinfo {year} {2010})}\BibitemShut {NoStop}%
\bibitem [{\citenamefont {Leroux}\ \emph {et~al.}(2010)\citenamefont {Leroux},
  \citenamefont {Schleier-Smith},\ and\ \citenamefont
  {Vuleti\ifmmode~\acute{c}\else \'{c}\fi{}}}]{LSM10}%
  \BibitemOpen
  \bibfield  {author} {\bibinfo {author} {\bibfnamefont {I.~D.}\ \bibnamefont
  {Leroux}}, \bibinfo {author} {\bibfnamefont {M.~H.}\ \bibnamefont
  {Schleier-Smith}}, \ and\ \bibinfo {author} {\bibfnamefont {V.}~\bibnamefont
  {Vuleti\ifmmode~\acute{c}\else \'{c}\fi{}}},\ }\href {\doibase
  10.1103/PhysRevLett.104.073602} {\bibfield  {journal} {\bibinfo  {journal}
  {Phys. Rev. Lett.}\ }\textbf {\bibinfo {volume} {104}},\ \bibinfo {pages}
  {073602} (\bibinfo {year} {2010})}\BibitemShut {NoStop}%
\bibitem [{\citenamefont {Chen}\ \emph {et~al.}(2011)\citenamefont {Chen},
  \citenamefont {Bohnet}, \citenamefont {Sankar}, \citenamefont {Dai},\ and\
  \citenamefont {Thompson}}]{CBS11}%
  \BibitemOpen
  \bibfield  {author} {\bibinfo {author} {\bibfnamefont {Z.}~\bibnamefont
  {Chen}}, \bibinfo {author} {\bibfnamefont {J.~G.}\ \bibnamefont {Bohnet}},
  \bibinfo {author} {\bibfnamefont {S.~R.}\ \bibnamefont {Sankar}}, \bibinfo
  {author} {\bibfnamefont {J.}~\bibnamefont {Dai}}, \ and\ \bibinfo {author}
  {\bibfnamefont {J.~K.}\ \bibnamefont {Thompson}},\ }\href {\doibase
  10.1103/PhysRevLett.106.133601} {\bibfield  {journal} {\bibinfo  {journal}
  {Phys. Rev. Lett.}\ }\textbf {\bibinfo {volume} {106}},\ \bibinfo {pages}
  {133601} (\bibinfo {year} {2011})}\BibitemShut {NoStop}%
\bibitem [{\citenamefont {Gustavsson}\ \emph {et~al.}(2008)\citenamefont
  {Gustavsson}, \citenamefont {Haller}, \citenamefont {Mark}, \citenamefont
  {Danzl}, \citenamefont {Rojas-Kopeinig},\ and\ \citenamefont
  {N\"agerl}}]{GHM08}%
  \BibitemOpen
  \bibfield  {author} {\bibinfo {author} {\bibfnamefont {M.}~\bibnamefont
  {Gustavsson}}, \bibinfo {author} {\bibfnamefont {E.}~\bibnamefont {Haller}},
  \bibinfo {author} {\bibfnamefont {M.~J.}\ \bibnamefont {Mark}}, \bibinfo
  {author} {\bibfnamefont {J.~G.}\ \bibnamefont {Danzl}}, \bibinfo {author}
  {\bibfnamefont {G.}~\bibnamefont {Rojas-Kopeinig}}, \ and\ \bibinfo {author}
  {\bibfnamefont {H.-C.}\ \bibnamefont {N\"agerl}},\ }\href {\doibase
  10.1103/PhysRevLett.100.080404} {\bibfield  {journal} {\bibinfo  {journal}
  {Phys. Rev. Lett.}\ }\textbf {\bibinfo {volume} {100}},\ \bibinfo {pages}
  {080404} (\bibinfo {year} {2008})}\BibitemShut {NoStop}%
\bibitem [{\citenamefont {Gross}\ \emph {et~al.}(2010)\citenamefont {Gross},
  \citenamefont {Zibold}, \citenamefont {Nicklas}, \citenamefont {Estève},\
  and\ \citenamefont {Oberthaler}}]{GZN10}%
  \BibitemOpen
  \bibfield  {author} {\bibinfo {author} {\bibfnamefont {C.}~\bibnamefont
  {Gross}}, \bibinfo {author} {\bibfnamefont {T.}~\bibnamefont {Zibold}},
  \bibinfo {author} {\bibfnamefont {E.}~\bibnamefont {Nicklas}}, \bibinfo
  {author} {\bibfnamefont {J.}~\bibnamefont {Estève}}, \ and\ \bibinfo
  {author} {\bibfnamefont {M.~K.}\ \bibnamefont {Oberthaler}},\ }\href
  {\doibase 10.1038/nature08919} {\bibfield  {journal} {\bibinfo  {journal}
  {Nature}\ }\textbf {\bibinfo {volume} {464}},\ \bibinfo {pages} {1165}
  (\bibinfo {year} {2010})}\BibitemShut {NoStop}%
\bibitem [{\citenamefont {Deutsch}\ \emph {et~al.}(2010)\citenamefont
  {Deutsch}, \citenamefont {Ramirez-Martinez}, \citenamefont {Lacro\^ute},
  \citenamefont {Reinhard}, \citenamefont {Schneider}, \citenamefont {Fuchs},
  \citenamefont {Pi\'echon}, \citenamefont {Lalo\"e}, \citenamefont {Reichel},\
  and\ \citenamefont {Rosenbusch}}]{DRL10}%
  \BibitemOpen
  \bibfield  {author} {\bibinfo {author} {\bibfnamefont {C.}~\bibnamefont
  {Deutsch}}, \bibinfo {author} {\bibfnamefont {F.}~\bibnamefont
  {Ramirez-Martinez}}, \bibinfo {author} {\bibfnamefont {C.}~\bibnamefont
  {Lacro\^ute}}, \bibinfo {author} {\bibfnamefont {F.}~\bibnamefont
  {Reinhard}}, \bibinfo {author} {\bibfnamefont {T.}~\bibnamefont {Schneider}},
  \bibinfo {author} {\bibfnamefont {J.~N.}\ \bibnamefont {Fuchs}}, \bibinfo
  {author} {\bibfnamefont {F.}~\bibnamefont {Pi\'echon}}, \bibinfo {author}
  {\bibfnamefont {F.}~\bibnamefont {Lalo\"e}}, \bibinfo {author} {\bibfnamefont
  {J.}~\bibnamefont {Reichel}}, \ and\ \bibinfo {author} {\bibfnamefont
  {P.}~\bibnamefont {Rosenbusch}},\ }\href {\doibase
  10.1103/PhysRevLett.105.020401} {\bibfield  {journal} {\bibinfo  {journal}
  {Phys. Rev. Lett.}\ }\textbf {\bibinfo {volume} {105}},\ \bibinfo {pages}
  {020401} (\bibinfo {year} {2010})}\BibitemShut {NoStop}%
\bibitem [{\citenamefont {Schawlow}\ and\ \citenamefont
  {Townes}(1958)}]{SCT58}%
  \BibitemOpen
  \bibfield  {author} {\bibinfo {author} {\bibfnamefont {A.~L.}\ \bibnamefont
  {Schawlow}}\ and\ \bibinfo {author} {\bibfnamefont {C.~H.}\ \bibnamefont
  {Townes}},\ }\href {\doibase 10.1103/PhysRev.112.1940} {\bibfield  {journal}
  {\bibinfo  {journal} {Phys. Rev.}\ }\textbf {\bibinfo {volume} {112}},\
  \bibinfo {pages} {1940} (\bibinfo {year} {1958})}\BibitemShut {NoStop}%
\bibitem [{\citenamefont {Goldenberg}\ \emph {et~al.}(1960)\citenamefont
  {Goldenberg}, \citenamefont {Kleppner},\ and\ \citenamefont
  {Ramsey}}]{GKR60}%
  \BibitemOpen
  \bibfield  {author} {\bibinfo {author} {\bibfnamefont {H.~M.}\ \bibnamefont
  {Goldenberg}}, \bibinfo {author} {\bibfnamefont {D.}~\bibnamefont
  {Kleppner}}, \ and\ \bibinfo {author} {\bibfnamefont {N.~F.}\ \bibnamefont
  {Ramsey}},\ }\href {\doibase 10.1103/PhysRevLett.5.361} {\bibfield  {journal}
  {\bibinfo  {journal} {Phys. Rev. Lett.}\ }\textbf {\bibinfo {volume} {5}},\
  \bibinfo {pages} {361} (\bibinfo {year} {1960})}\BibitemShut {NoStop}%
\bibitem [{\citenamefont {Meiser}\ \emph {et~al.}(2009)\citenamefont {Meiser},
  \citenamefont {Ye}, \citenamefont {Carlson},\ and\ \citenamefont
  {Holland}}]{MYC09}%
  \BibitemOpen
  \bibfield  {author} {\bibinfo {author} {\bibfnamefont {D.}~\bibnamefont
  {Meiser}}, \bibinfo {author} {\bibfnamefont {J.}~\bibnamefont {Ye}}, \bibinfo
  {author} {\bibfnamefont {D.~R.}\ \bibnamefont {Carlson}}, \ and\ \bibinfo
  {author} {\bibfnamefont {M.~J.}\ \bibnamefont {Holland}},\ }\href {\doibase
  10.1103/PhysRevLett.102.163601} {\bibfield  {journal} {\bibinfo  {journal}
  {Phys. Rev. Lett.}\ }\textbf {\bibinfo {volume} {102}},\ \bibinfo {pages}
  {163601} (\bibinfo {year} {2009})}\BibitemShut {NoStop}%
\bibitem [{\citenamefont {Bohnet}\ \emph {et~al.}(2012)\citenamefont {Bohnet},
  \citenamefont {Chen}, \citenamefont {Weiner}, \citenamefont {Meiser},
  \citenamefont {Holland},\ and\ \citenamefont {Thompson}}]{BCW12}%
  \BibitemOpen
  \bibfield  {author} {\bibinfo {author} {\bibfnamefont {J.~G.}\ \bibnamefont
  {Bohnet}}, \bibinfo {author} {\bibfnamefont {Z.}~\bibnamefont {Chen}},
  \bibinfo {author} {\bibfnamefont {J.~M.}\ \bibnamefont {Weiner}}, \bibinfo
  {author} {\bibfnamefont {D.}~\bibnamefont {Meiser}}, \bibinfo {author}
  {\bibfnamefont {M.~J.}\ \bibnamefont {Holland}}, \ and\ \bibinfo {author}
  {\bibfnamefont {J.~K.}\ \bibnamefont {Thompson}},\ }\href@noop {} {\bibfield
  {journal} {\bibinfo  {journal} {Nature}\ }\textbf {\bibinfo {volume} {484}},\
  \bibinfo {pages} {78} (\bibinfo {year} {2012})}\BibitemShut {NoStop}%
\bibitem [{\citenamefont {Feynman}\ \emph {et~al.}(1957)\citenamefont
  {Feynman}, \citenamefont {Vernon},\ and\ \citenamefont {Hellwarth}}]{FVH57}%
  \BibitemOpen
  \bibfield  {author} {\bibinfo {author} {\bibfnamefont {R.~P.}\ \bibnamefont
  {Feynman}}, \bibinfo {author} {\bibfnamefont {F.~L.}\ \bibnamefont {Vernon}},
  \ and\ \bibinfo {author} {\bibfnamefont {R.~W.}\ \bibnamefont {Hellwarth}},\
  }\href {\doibase 10.1063/1.1722572} {\bibfield  {journal} {\bibinfo
  {journal} {Journal of Applied Physics}\ }\textbf {\bibinfo {volume} {28}},\
  \bibinfo {pages} {49 } (\bibinfo {year} {1957})}\BibitemShut {NoStop}%
\bibitem [{\citenamefont {Dos~Santos}\ \emph
  {et~al.}(2002)\citenamefont {Dos~Santos}, \citenamefont {Marion},
  \citenamefont {Bize}, \citenamefont {Sortais}, \citenamefont {Clairon},\ and\
  \citenamefont {Salomon}}]{Saloman02}%
  \BibitemOpen
  \bibfield  {author} {\bibinfo {author} {\bibfnamefont {F. P.}~\bibnamefont
  {Dos~Santos}}, \bibinfo {author} {\bibfnamefont {H.}~\bibnamefont
  {Marion}}, \bibinfo {author} {\bibfnamefont {S.}~\bibnamefont {Bize}},
  \bibinfo {author} {\bibfnamefont {Y.}~\bibnamefont {Sortais}}, \bibinfo
  {author} {\bibfnamefont {A.}~\bibnamefont {Clairon}}, \ and\ \bibinfo
  {author} {\bibfnamefont {C.}~\bibnamefont {Salomon}},\ }\href {\doibase
  10.1103/PhysRevLett.89.233004} {\bibfield  {journal} {\bibinfo  {journal}
  {Phys. Rev. Lett.}\ }\textbf {\bibinfo {volume} {89}},\ \bibinfo {pages}
  {233004} (\bibinfo {year} {2002})}\BibitemShut {NoStop}%
\bibitem [{\citenamefont {Peters}\ \emph {et~al.}(2001)\citenamefont {Peters},
  \citenamefont {Chung},\ and\ \citenamefont {Chu}}]{PCC01}%
  \BibitemOpen
  \bibfield  {author} {\bibinfo {author} {\bibfnamefont {A.}~\bibnamefont
  {Peters}}, \bibinfo {author} {\bibfnamefont {K.~Y.}\ \bibnamefont {Chung}}, \
  and\ \bibinfo {author} {\bibfnamefont {S.}~\bibnamefont {Chu}},\ }\href
  {http://stacks.iop.org/0026-1394/38/i=1/a=4} {\bibfield  {journal} {\bibinfo
  {journal} {Metrologia}\ }\textbf {\bibinfo {volume} {38}},\ \bibinfo {pages}
  {25} (\bibinfo {year} {2001})}\BibitemShut {NoStop}%
\bibitem [{\citenamefont {Biedermann}\ \emph {et~al.}(2009)\citenamefont
  {Biedermann}, \citenamefont {Wu}, \citenamefont {Deslauriers}, \citenamefont
  {Takase},\ and\ \citenamefont {Kasevich}}]{Biedermann09}%
  \BibitemOpen
  \bibfield  {author} {\bibinfo {author} {\bibfnamefont {G.~W.}\ \bibnamefont
  {Biedermann}}, \bibinfo {author} {\bibfnamefont {X.}~\bibnamefont {Wu}},
  \bibinfo {author} {\bibfnamefont {L.}~\bibnamefont {Deslauriers}}, \bibinfo
  {author} {\bibfnamefont {K.}~\bibnamefont {Takase}}, \ and\ \bibinfo {author}
  {\bibfnamefont {M.~A.}\ \bibnamefont {Kasevich}},\ }\href {\doibase
  10.1364/OL.34.000347} {\bibfield  {journal} {\bibinfo  {journal} {Opt.
  Lett.}\ }\textbf {\bibinfo {volume} {34}},\ \bibinfo {pages} {347} (\bibinfo
  {year} {2009})}\BibitemShut {NoStop}%
\bibitem [{\citenamefont {Ramsey}(1950)}]{PhysRev.78.695}%
  \BibitemOpen
  \bibfield  {author} {\bibinfo {author} {\bibfnamefont {N.~F.}\ \bibnamefont
  {Ramsey}},\ }\href {\doibase 10.1103/PhysRev.78.695} {\bibfield  {journal}
  {\bibinfo  {journal} {Phys. Rev.}\ }\textbf {\bibinfo {volume} {78}},\
  \bibinfo {pages} {695} (\bibinfo {year} {1950})}\BibitemShut {NoStop}%
\bibitem [{\citenamefont {Katori}(2011)}]{Katori2011}%
  \BibitemOpen
  \bibfield  {author} {\bibinfo {author} {\bibfnamefont {H.}~\bibnamefont
  {Katori}},\ }\href {\doibase 10.1038/nphoton.2011.45} {\bibfield  {journal}
  {\bibinfo  {journal} {Nature Photonics}\ }\textbf {\bibinfo {volume} {5}},\
  \bibinfo {pages} {203} (\bibinfo {year} {2011})}\BibitemShut {NoStop}%
\bibitem [{\citenamefont {Leibrandt}\ \emph {et~al.}(2011)\citenamefont
  {Leibrandt}, \citenamefont {Thorpe}, \citenamefont {Bergquist},\ and\
  \citenamefont {Rosenband}}]{LTB11}%
  \BibitemOpen
  \bibfield  {author} {\bibinfo {author} {\bibfnamefont {D.~R.}\ \bibnamefont
  {Leibrandt}}, \bibinfo {author} {\bibfnamefont {M.~J.}\ \bibnamefont
  {Thorpe}}, \bibinfo {author} {\bibfnamefont {J.~C.}\ \bibnamefont
  {Bergquist}}, \ and\ \bibinfo {author} {\bibfnamefont {T.}~\bibnamefont
  {Rosenband}},\ }\href {\doibase 10.1364/OE.19.010278} {\bibfield  {journal}
  {\bibinfo  {journal} {Opt. Express}\ }\textbf {\bibinfo {volume} {19}},\
  \bibinfo {pages} {10278} (\bibinfo {year} {2011})}\BibitemShut {NoStop}%
\bibitem [{\citenamefont {Brennecke}\ \emph {et~al.}(2007)\citenamefont
  {Brennecke}, \citenamefont {Donner}, \citenamefont {Ritter}, \citenamefont
  {Bourdel}, \citenamefont {Köhl},\ and\ \citenamefont {Esslinger}}]{BDR07}%
  \BibitemOpen
  \bibfield  {author} {\bibinfo {author} {\bibfnamefont {F.}~\bibnamefont
  {Brennecke}}, \bibinfo {author} {\bibfnamefont {T.}~\bibnamefont {Donner}},
  \bibinfo {author} {\bibfnamefont {S.}~\bibnamefont {Ritter}}, \bibinfo
  {author} {\bibfnamefont {T.}~\bibnamefont {Bourdel}}, \bibinfo {author}
  {\bibfnamefont {M.}~\bibnamefont {Köhl}}, \ and\ \bibinfo {author}
  {\bibfnamefont {T.}~\bibnamefont {Esslinger}},\ }\href {\doibase
  10.1038/nature06120} {\bibfield  {journal} {\bibinfo  {journal} {Nature}\
  }\textbf {\bibinfo {volume} {450}},\ \bibinfo {pages} {268} (\bibinfo {year}
  {2007})}\BibitemShut {NoStop}%
\bibitem [{\citenamefont {Tanji}\ \emph {et~al.}(2009)\citenamefont {Tanji},
  \citenamefont {Ghosh}, \citenamefont {Simon}, \citenamefont {Bloom},\ and\
  \citenamefont {Vuleti\ifmmode~\acute{c}\else \'{c}\fi{}}}]{TGS09}%
  \BibitemOpen
  \bibfield  {author} {\bibinfo {author} {\bibfnamefont {H.}~\bibnamefont
  {Tanji}}, \bibinfo {author} {\bibfnamefont {S.}~\bibnamefont {Ghosh}},
  \bibinfo {author} {\bibfnamefont {J.}~\bibnamefont {Simon}}, \bibinfo
  {author} {\bibfnamefont {B.}~\bibnamefont {Bloom}}, \ and\ \bibinfo {author}
  {\bibfnamefont {V.}~\bibnamefont {Vuleti\ifmmode~\acute{c}\else
  \'{c}\fi{}}},\ }\href {\doibase 10.1103/PhysRevLett.103.043601} {\bibfield
  {journal} {\bibinfo  {journal} {Phys. Rev. Lett.}\ }\textbf {\bibinfo
  {volume} {103}},\ \bibinfo {pages} {043601} (\bibinfo {year}
  {2009})}\BibitemShut {NoStop}%
\bibitem [{\citenamefont {Mekhov}\ and\ \citenamefont {Ritsch}(2009)}]{MR09}%
  \BibitemOpen
  \bibfield  {author} {\bibinfo {author} {\bibfnamefont {I.~B.}\ \bibnamefont
  {Mekhov}}\ and\ \bibinfo {author} {\bibfnamefont {H.}~\bibnamefont
  {Ritsch}},\ }\href {\doibase 10.1103/PhysRevLett.102.020403} {\bibfield
  {journal} {\bibinfo  {journal} {Phys. Rev. Lett.}\ }\textbf {\bibinfo
  {volume} {102}},\ \bibinfo {pages} {020403} (\bibinfo {year}
  {2009})}\BibitemShut {NoStop}%
\bibitem [{\citenamefont {Tanji-Suzuki}\ \emph {et~al.}(2011)\citenamefont
  {Tanji-Suzuki}, \citenamefont {Chen}, \citenamefont {Landig}, \citenamefont
  {Simon},\ and\ \citenamefont {Vuleti\ifmmode~\acute{c}\else
  \'{c}\fi{}}}]{TWL11}%
  \BibitemOpen
  \bibfield  {author} {\bibinfo {author} {\bibfnamefont {H.}~\bibnamefont
  {Tanji-Suzuki}}, \bibinfo {author} {\bibfnamefont {W.}~\bibnamefont {Chen}},
  \bibinfo {author} {\bibfnamefont {R.}~\bibnamefont {Landig}}, \bibinfo
  {author} {\bibfnamefont {J.}~\bibnamefont {Simon}}, \ and\ \bibinfo {author}
  {\bibfnamefont {V.}~\bibnamefont {Vuleti\ifmmode~\acute{c}\else
  \'{c}\fi{}}},\ }\href {\doibase 10.1126/science.1208066} {\bibfield
  {journal} {\bibinfo  {journal} {Science}\ }\textbf {\bibinfo {volume}
  {333}},\ \bibinfo {pages} {1266} (\bibinfo {year} {2011})}\BibitemShut
  {NoStop}%
\bibitem [{\citenamefont {Dayan}\ \emph {et~al.}(2008)\citenamefont {Dayan},
  \citenamefont {Parkins}, \citenamefont {Aoki}, \citenamefont {Ostby},
  \citenamefont {Vahala},\ and\ \citenamefont {Kimble}}]{Dayan08}%
  \BibitemOpen
  \bibfield  {author} {\bibinfo {author} {\bibfnamefont {B.}~\bibnamefont
  {Dayan}}, \bibinfo {author} {\bibfnamefont {A.~S.}\ \bibnamefont {Parkins}},
  \bibinfo {author} {\bibfnamefont {T.}~\bibnamefont {Aoki}}, \bibinfo {author}
  {\bibfnamefont {E.~P.}\ \bibnamefont {Ostby}}, \bibinfo {author}
  {\bibfnamefont {K.~J.}\ \bibnamefont {Vahala}}, \ and\ \bibinfo {author}
  {\bibfnamefont {H.~J.}\ \bibnamefont {Kimble}},\ }\href {\doibase
  10.1126/science.1152261} {\bibfield  {journal} {\bibinfo  {journal}
  {Science}\ }\textbf {\bibinfo {volume} {319}},\ \bibinfo {pages} {1062}
  (\bibinfo {year} {2008})}\BibitemShut {NoStop}%
\bibitem [{\citenamefont {Specht}\ \emph {et~al.}(2011)\citenamefont {Specht},
  \citenamefont {Nölleke}, \citenamefont {Reiserer}, \citenamefont {Uphoff},
  \citenamefont {Figueroa}, \citenamefont {Ritter},\ and\ \citenamefont
  {Rempe}}]{SNR11}%
  \BibitemOpen
  \bibfield  {author} {\bibinfo {author} {\bibfnamefont {H.~P.}\ \bibnamefont
  {Specht}}, \bibinfo {author} {\bibfnamefont {C.}~\bibnamefont {Nölleke}},
  \bibinfo {author} {\bibfnamefont {A.}~\bibnamefont {Reiserer}}, \bibinfo
  {author} {\bibfnamefont {M.}~\bibnamefont {Uphoff}}, \bibinfo {author}
  {\bibfnamefont {E.}~\bibnamefont {Figueroa}}, \bibinfo {author}
  {\bibfnamefont {S.}~\bibnamefont {Ritter}}, \ and\ \bibinfo {author}
  {\bibfnamefont {G.}~\bibnamefont {Rempe}},\ }\href {\doibase
  10.1038/nature09997} {\bibfield  {journal} {\bibinfo  {journal} {Nature}\
  }\textbf {\bibinfo {volume} {473}},\ \bibinfo {pages} {190–193} (\bibinfo
  {year} {2011})}\BibitemShut {NoStop}%
\bibitem [{\citenamefont {Lodewyck}\ \emph {et~al.}(2009)\citenamefont
  {Lodewyck}, \citenamefont {Westergaard},\ and\ \citenamefont
  {Lemonde}}]{LWL09}%
  \BibitemOpen
  \bibfield  {author} {\bibinfo {author} {\bibfnamefont {J.}~\bibnamefont
  {Lodewyck}}, \bibinfo {author} {\bibfnamefont {P.~G.}\ \bibnamefont
  {Westergaard}}, \ and\ \bibinfo {author} {\bibfnamefont {P.}~\bibnamefont
  {Lemonde}},\ }\href {\doibase 10.1103/PhysRevA.79.061401} {\bibfield
  {journal} {\bibinfo  {journal} {Phys. Rev. A}\ }\textbf {\bibinfo {volume}
  {79}},\ \bibinfo {pages} {061401} (\bibinfo {year} {2009})}\BibitemShut
  {NoStop}%
\bibitem [{\citenamefont {Westergaard}\ \emph {et~al.}(2010)\citenamefont
  {Westergaard}, \citenamefont {Lodewyck},\ and\ \citenamefont
  {Lemonde}}]{Westergaard2010}%
  \BibitemOpen
  \bibfield  {author} {\bibinfo {author} {\bibfnamefont {P.}~\bibnamefont
  {Westergaard}}, \bibinfo {author} {\bibfnamefont {J.}~\bibnamefont
  {Lodewyck}}, \ and\ \bibinfo {author} {\bibfnamefont {P.}~\bibnamefont
  {Lemonde}},\ }\href {\doibase 10.1109/TUFFC.2010.1457} {\bibfield  {journal}
  {\bibinfo  {journal} {IEEE Trans. Ultr. Ferr. Freq. Cont.}\ }\textbf
  {\bibinfo {volume} {57}},\ \bibinfo {pages} {623} (\bibinfo {year}
  {2010})}\BibitemShut {NoStop}%
\bibitem [{\citenamefont {Duan}\ \emph {et~al.}(2001)\citenamefont {Duan},
  \citenamefont {Lukin}, \citenamefont {Cirac},\ and\ \citenamefont
  {Zoller}}]{DLCZ01}%
  \BibitemOpen
  \bibfield  {author} {\bibinfo {author} {\bibfnamefont {L.-M.}\ \bibnamefont
  {Duan}}, \bibinfo {author} {\bibfnamefont {M.~D.}\ \bibnamefont {Lukin}},
  \bibinfo {author} {\bibfnamefont {J.~I.}\ \bibnamefont {Cirac}}, \ and\
  \bibinfo {author} {\bibfnamefont {P.}~\bibnamefont {Zoller}},\ }\href
  {\doibase 10.1038/35106500} {\bibfield  {journal} {\bibinfo  {journal}
  {Nature}\ }\textbf {\bibinfo {volume} {414}},\ \bibinfo {pages} {413}
  (\bibinfo {year} {2001})}\BibitemShut {NoStop}%
\bibitem [{\citenamefont {Boozer}\ \emph {et~al.}(2007)\citenamefont {Boozer},
  \citenamefont {Boca}, \citenamefont {Miller}, \citenamefont {Northup},\ and\
  \citenamefont {Kimble}}]{BBM07}%
  \BibitemOpen
  \bibfield  {author} {\bibinfo {author} {\bibfnamefont {A.~D.}\ \bibnamefont
  {Boozer}}, \bibinfo {author} {\bibfnamefont {A.}~\bibnamefont {Boca}},
  \bibinfo {author} {\bibfnamefont {R.}~\bibnamefont {Miller}}, \bibinfo
  {author} {\bibfnamefont {T.~E.}\ \bibnamefont {Northup}}, \ and\ \bibinfo
  {author} {\bibfnamefont {H.~J.}\ \bibnamefont {Kimble}},\ }\href {\doibase
  10.1103/PhysRevLett.98.193601} {\bibfield  {journal} {\bibinfo  {journal}
  {Phys. Rev. Lett.}\ }\textbf {\bibinfo {volume} {98}},\ \bibinfo {pages}
  {193601} (\bibinfo {year} {2007})}\BibitemShut {NoStop}%
\bibitem [{\citenamefont {Choi}\ \emph {et~al.}(2010)\citenamefont {Choi},
  \citenamefont {Goban}, \citenamefont {Papp}, \citenamefont {van Enk},\ and\
  \citenamefont {Kimble}}]{CGP10}%
  \BibitemOpen
  \bibfield  {author} {\bibinfo {author} {\bibfnamefont {K.~S.}\ \bibnamefont
  {Choi}}, \bibinfo {author} {\bibfnamefont {A.}~\bibnamefont {Goban}},
  \bibinfo {author} {\bibfnamefont {S.~B.}\ \bibnamefont {Papp}}, \bibinfo
  {author} {\bibfnamefont {S.~J.}\ \bibnamefont {van Enk}}, \ and\ \bibinfo
  {author} {\bibfnamefont {H.~J.}\ \bibnamefont {Kimble}},\ }\href {\doibase
  10.1038/nature09568} {\bibfield  {journal} {\bibinfo  {journal} {Nature}\
  }\textbf {\bibinfo {volume} {468}},\ \bibinfo {pages} {412} (\bibinfo {year}
  {2010})}\BibitemShut {NoStop}%
\bibitem [{\citenamefont {Radnaev}\ \emph {et~al.}(2010)\citenamefont
  {Radnaev}, \citenamefont {Dudin}, \citenamefont {Zhao}, \citenamefont {Jen},
  \citenamefont {Jenkins}, \citenamefont {Kuzmich},\ and\ \citenamefont
  {Kennedy}}]{RDZ10}%
  \BibitemOpen
  \bibfield  {author} {\bibinfo {author} {\bibfnamefont {A.~G.}\ \bibnamefont
  {Radnaev}}, \bibinfo {author} {\bibfnamefont {Y.~O.}\ \bibnamefont {Dudin}},
  \bibinfo {author} {\bibfnamefont {R.}~\bibnamefont {Zhao}}, \bibinfo {author}
  {\bibfnamefont {H.~H.}\ \bibnamefont {Jen}}, \bibinfo {author} {\bibfnamefont
  {S.~D.}\ \bibnamefont {Jenkins}}, \bibinfo {author} {\bibfnamefont
  {A.}~\bibnamefont {Kuzmich}}, \ and\ \bibinfo {author} {\bibfnamefont
  {T.~A.~B.}\ \bibnamefont {Kennedy}},\ }\href {\doibase 10.1038/nphys1773}
  {\bibfield  {journal} {\bibinfo  {journal} {Nat Phys}\ }\textbf {\bibinfo
  {volume} {6}},\ \bibinfo {pages} {894} (\bibinfo {year} {2010})}\BibitemShut
  {NoStop}%
\bibitem [{\citenamefont {Lettner}\ \emph {et~al.}(2011)\citenamefont
  {Lettner}, \citenamefont {M\"ucke}, \citenamefont {Riedl}, \citenamefont
  {Vo}, \citenamefont {Hahn}, \citenamefont {Baur}, \citenamefont {Bochmann},
  \citenamefont {Ritter}, \citenamefont {D\"urr},\ and\ \citenamefont
  {Rempe}}]{LMR11}%
  \BibitemOpen
  \bibfield  {author} {\bibinfo {author} {\bibfnamefont {M.}~\bibnamefont
  {Lettner}}, \bibinfo {author} {\bibfnamefont {M.}~\bibnamefont {M\"ucke}},
  \bibinfo {author} {\bibfnamefont {S.}~\bibnamefont {Riedl}}, \bibinfo
  {author} {\bibfnamefont {C.}~\bibnamefont {Vo}}, \bibinfo {author}
  {\bibfnamefont {C.}~\bibnamefont {Hahn}}, \bibinfo {author} {\bibfnamefont
  {S.}~\bibnamefont {Baur}}, \bibinfo {author} {\bibfnamefont {J.}~\bibnamefont
  {Bochmann}}, \bibinfo {author} {\bibfnamefont {S.}~\bibnamefont {Ritter}},
  \bibinfo {author} {\bibfnamefont {S.}~\bibnamefont {D\"urr}}, \ and\ \bibinfo
  {author} {\bibfnamefont {G.}~\bibnamefont {Rempe}},\ }\href {\doibase
  10.1103/PhysRevLett.106.210503} {\bibfield  {journal} {\bibinfo  {journal}
  {Phys. Rev. Lett.}\ }\textbf {\bibinfo {volume} {106}},\ \bibinfo {pages}
  {210503} (\bibinfo {year} {2011})}\BibitemShut {NoStop}%
\bibitem [{\citenamefont {Kuppens}\ \emph {et~al.}(1994)\citenamefont
  {Kuppens}, \citenamefont {van Exter},\ and\ \citenamefont
  {Woerdman}}]{PhysRevLett.72.3815}%
  \BibitemOpen
  \bibfield  {author} {\bibinfo {author} {\bibfnamefont {S.~J.~M.}\
  \bibnamefont {Kuppens}}, \bibinfo {author} {\bibfnamefont {M.~P.}\
  \bibnamefont {van Exter}}, \ and\ \bibinfo {author} {\bibfnamefont {J.~P.}\
  \bibnamefont {Woerdman}},\ }\href {\doibase 10.1103/PhysRevLett.72.3815}
  {\bibfield  {journal} {\bibinfo  {journal} {Phys. Rev. Lett.}\ }\textbf
  {\bibinfo {volume} {72}},\ \bibinfo {pages} {3815} (\bibinfo {year}
  {1994})}\BibitemShut {NoStop}%
\bibitem [{\citenamefont {Kolobov}\ \emph {et~al.}(1993)\citenamefont
  {Kolobov}, \citenamefont {Davidovich}, \citenamefont {Giacobino},\ and\
  \citenamefont {Fabre}}]{PhysRevA.47.1431}%
  \BibitemOpen
  \bibfield  {author} {\bibinfo {author} {\bibfnamefont {M.~I.}\ \bibnamefont
  {Kolobov}}, \bibinfo {author} {\bibfnamefont {L.}~\bibnamefont {Davidovich}},
  \bibinfo {author} {\bibfnamefont {E.}~\bibnamefont {Giacobino}}, \ and\
  \bibinfo {author} {\bibfnamefont {C.}~\bibnamefont {Fabre}},\ }\href
  {\doibase 10.1103/PhysRevA.47.1431} {\bibfield  {journal} {\bibinfo
  {journal} {Phys. Rev. A}\ }\textbf {\bibinfo {volume} {47}},\ \bibinfo
  {pages} {1431} (\bibinfo {year} {1993})}\BibitemShut {NoStop}%
\bibitem [{\citenamefont {Kitching}\ \emph {et~al.}(2011)\citenamefont
  {Kitching}, \citenamefont {Knappe},\ and\ \citenamefont
  {Donley}}]{Kitching2011}%
  \BibitemOpen
  \bibfield  {author} {\bibinfo {author} {\bibfnamefont {J.}~\bibnamefont
  {Kitching}}, \bibinfo {author} {\bibfnamefont {S.}~\bibnamefont {Knappe}}, \
  and\ \bibinfo {author} {\bibfnamefont {E.~A.}\ \bibnamefont {Donley}},\
  }\href {\doibase 10.1109/JSEN.2011.2157679} {\bibfield  {journal} {\bibinfo
  {journal} {IEEE Sensors Journal}\ }\textbf {\bibinfo {volume} {11}},\
  \bibinfo {pages} {1749} (\bibinfo {year} {2011})}\BibitemShut {NoStop}%
\bibitem [{\citenamefont {Gross}\ and\ \citenamefont
  {Haroche}(1982)}]{Gross1982}%
  \BibitemOpen
  \bibfield  {author} {\bibinfo {author} {\bibfnamefont {M.}~\bibnamefont
  {Gross}}\ and\ \bibinfo {author} {\bibfnamefont {S.}~\bibnamefont
  {Haroche}},\ }\href {\doibase 10.1016/0370-1573(82)90102-8} {\bibfield
  {journal} {\bibinfo  {journal} {Physics Reports}\ }\textbf {\bibinfo {volume}
  {93}},\ \bibinfo {pages} {301 } (\bibinfo {year} {1982})}\BibitemShut
  {NoStop}%
\bibitem [{\citenamefont {Kalman}\ and\ \citenamefont
  {Bucy}(1961)}]{Kalman1961}%
  \BibitemOpen
  \bibfield  {author} {\bibinfo {author} {\bibfnamefont {R.~E.}\ \bibnamefont
  {Kalman}}\ and\ \bibinfo {author} {\bibfnamefont {R.~S.}\ \bibnamefont
  {Bucy}},\ }\href@noop {} {\bibfield  {journal} {\bibinfo  {journal} {Journal
  of Basic Engineering}\ }\textbf {\bibinfo {volume} {83}},\ \bibinfo {pages}
  {95} (\bibinfo {year} {1961})}\BibitemShut {NoStop}%
\bibitem [{SOM()}]{SOM}%
  \BibitemOpen
  \href@noop {} {\enquote {\bibinfo {title} {See supporting online material},}\
  }\BibitemShut {NoStop}%
\bibitem [{\citenamefont {Barber}\ \emph {et~al.}(2006)\citenamefont {Barber},
  \citenamefont {Hoyt}, \citenamefont {Oates}, \citenamefont {Hollberg},
  \citenamefont {Taichenachev},\ and\ \citenamefont {Yudin}}]{BHO06}%
  \BibitemOpen
  \bibfield  {author} {\bibinfo {author} {\bibfnamefont {Z.~W.}\ \bibnamefont
  {Barber}}, \bibinfo {author} {\bibfnamefont {C.~W.}\ \bibnamefont {Hoyt}},
  \bibinfo {author} {\bibfnamefont {C.~W.}\ \bibnamefont {Oates}}, \bibinfo
  {author} {\bibfnamefont {L.}~\bibnamefont {Hollberg}}, \bibinfo {author}
  {\bibfnamefont {A.~V.}\ \bibnamefont {Taichenachev}}, \ and\ \bibinfo
  {author} {\bibfnamefont {V.~I.}\ \bibnamefont {Yudin}},\ }\href {\doibase
  10.1103/PhysRevLett.96.083002} {\bibfield  {journal} {\bibinfo  {journal}
  {Phys. Rev. Lett.}\ }\textbf {\bibinfo {volume} {96}},\ \bibinfo {pages}
  {083002} (\bibinfo {year} {2006})}\BibitemShut {NoStop}%
\bibitem [{\citenamefont {Santra}\ \emph {et~al.}(2005)\citenamefont {Santra},
  \citenamefont {Arimondo}, \citenamefont {Ido}, \citenamefont {Greene},\ and\
  \citenamefont {Ye}}]{SAI05}%
  \BibitemOpen
  \bibfield  {author} {\bibinfo {author} {\bibfnamefont {R.}~\bibnamefont
  {Santra}}, \bibinfo {author} {\bibfnamefont {E.}~\bibnamefont {Arimondo}},
  \bibinfo {author} {\bibfnamefont {T.}~\bibnamefont {Ido}}, \bibinfo {author}
  {\bibfnamefont {C.~H.}\ \bibnamefont {Greene}}, \ and\ \bibinfo {author}
  {\bibfnamefont {J.}~\bibnamefont {Ye}},\ }\href {\doibase
  10.1103/PhysRevLett.94.173002} {\bibfield  {journal} {\bibinfo  {journal}
  {Phys. Rev. Lett.}\ }\textbf {\bibinfo {volume} {94}},\ \bibinfo {pages}
  {173002} (\bibinfo {year} {2005})}\BibitemShut {NoStop}%
\end{thebibliography}

\begin{thebibliography}{8}%
\makeatletter
\providecommand \@ifxundefined [1]{%
 \@ifx{#1\undefined}
}%
\providecommand \@ifnum [1]{%
 \ifnum #1\expandafter \@firstoftwo
 \else \expandafter \@secondoftwo
 \fi
}%
\providecommand \@ifx [1]{%
 \ifx #1\expandafter \@firstoftwo
 \else \expandafter \@secondoftwo
 \fi
}%
\providecommand \natexlab [1]{#1}%
\providecommand \enquote  [1]{``#1''}%
\providecommand \bibnamefont  [1]{#1}%
\providecommand \bibfnamefont [1]{#1}%
\providecommand \citenamefont [1]{#1}%
\providecommand \href@noop [0]{\@secondoftwo}%
\providecommand \href [0]{\begingroup \@sanitize@url \@href}%
\providecommand \@href[1]{\@@startlink{#1}\@@href}%
\providecommand \@@href[1]{\endgroup#1\@@endlink}%
\providecommand \@sanitize@url [0]{\catcode `\\12\catcode `\$12\catcode
  `\&12\catcode `\#12\catcode `\^12\catcode `\_12\catcode `\%12\relax}%
\providecommand \@@startlink[1]{}%
\providecommand \@@endlink[0]{}%
\providecommand \url  [0]{\begingroup\@sanitize@url \@url }%
\providecommand \@url [1]{\endgroup\@href {#1}{\urlprefix }}%
\providecommand \urlprefix  [0]{URL }%
\providecommand \Eprint [0]{\href }%
\providecommand \doibase [0]{http://dx.doi.org/}%
\providecommand \selectlanguage [0]{\@gobble}%
\providecommand \bibinfo  [0]{\@secondoftwo}%
\providecommand \bibfield  [0]{\@secondoftwo}%
\providecommand \translation [1]{[#1]}%
\providecommand \BibitemOpen [0]{}%
\providecommand \bibitemStop [0]{}%
\providecommand \bibitemNoStop [0]{.\EOS\space}%
\providecommand \EOS [0]{\spacefactor3000\relax}%
\providecommand \BibitemShut  [1]{\csname bibitem#1\endcsname}%
\let\auto@bib@innerbib\@empty
\bibitem [{\citenamefont {Kalman}(1960)}]{Kalman1960}%
  \BibitemOpen
  \bibfield  {author} {\bibinfo {author} {\bibfnamefont {R.~E.}\ \bibnamefont
  {Kalman}},\ }\href@noop {} {\bibfield  {journal} {\bibinfo  {journal}
  {Journal of Basic Engineering}\ }\textbf {\bibinfo {volume} {82}},\ \bibinfo
  {pages} {35} (\bibinfo {year} {1960})}\BibitemShut {NoStop}%
\bibitem [{\citenamefont {Kalman}\ and\ \citenamefont
  {Bucy}(1961)}]{Kalman1961}%
  \BibitemOpen
  \bibfield  {author} {\bibinfo {author} {\bibfnamefont {R.~E.}\ \bibnamefont
  {Kalman}}\ and\ \bibinfo {author} {\bibfnamefont {R.~S.}\ \bibnamefont
  {Bucy}},\ }\href@noop {} {\bibfield  {journal} {\bibinfo  {journal} {Journal
  of Basic Engineering}\ }\textbf {\bibinfo {volume} {83}},\ \bibinfo {pages}
  {95} (\bibinfo {year} {1961})}\BibitemShut {NoStop}%
\bibitem [{\citenamefont {Zarchan}\ and\ \citenamefont
  {Musoff}(2001)}]{zarchan_fundamentals_2001}%
  \BibitemOpen
  \bibfield  {author} {\bibinfo {author} {\bibfnamefont {P.}~\bibnamefont
  {Zarchan}}\ and\ \bibinfo {author} {\bibfnamefont {H.}~\bibnamefont
  {Musoff}},\ }\href@noop {} {\emph {\bibinfo {title} {Fundamentals of Kalman
  filtering: a Practical Approach}}}\ (\bibinfo  {publisher} {Amer Inst of
  Aeronautics \&},\ \bibinfo {year} {2001})\BibitemShut {NoStop}%
\bibitem [{\citenamefont {Meiser}\ \emph {et~al.}(2009)\citenamefont {Meiser},
  \citenamefont {Ye}, \citenamefont {Carlson},\ and\ \citenamefont
  {Holland}}]{MYC09}%
  \BibitemOpen
  \bibfield  {author} {\bibinfo {author} {\bibfnamefont {D.}~\bibnamefont
  {Meiser}}, \bibinfo {author} {\bibfnamefont {J.}~\bibnamefont {Ye}}, \bibinfo
  {author} {\bibfnamefont {D.~R.}\ \bibnamefont {Carlson}}, \ and\ \bibinfo
  {author} {\bibfnamefont {M.~J.}\ \bibnamefont {Holland}},\ }\href {\doibase
  10.1103/PhysRevLett.102.163601} {\bibfield  {journal} {\bibinfo  {journal}
  {Phys. Rev. Lett.}\ }\textbf {\bibinfo {volume} {102}},\ \bibinfo {pages}
  {163601} (\bibinfo {year} {2009})}\BibitemShut {NoStop}%
\bibitem [{\citenamefont {Bohnet}\ \emph {et~al.}(2012)\citenamefont {Bohnet},
  \citenamefont {Chen}, \citenamefont {Weiner}, \citenamefont {Meiser},
  \citenamefont {Holland},\ and\ \citenamefont {Thompson}}]{BCW12}%
  \BibitemOpen
  \bibfield  {author} {\bibinfo {author} {\bibfnamefont {J.~G.}\ \bibnamefont
  {Bohnet}}, \bibinfo {author} {\bibfnamefont {Z.}~\bibnamefont {Chen}},
  \bibinfo {author} {\bibfnamefont {J.~M.}\ \bibnamefont {Weiner}}, \bibinfo
  {author} {\bibfnamefont {D.}~\bibnamefont {Meiser}}, \bibinfo {author}
  {\bibfnamefont {M.~J.}\ \bibnamefont {Holland}}, \ and\ \bibinfo {author}
  {\bibfnamefont {J.~K.}\ \bibnamefont {Thompson}},\ }\href@noop {} {\bibfield
  {journal} {\bibinfo  {journal} {Nature}\ }\textbf {\bibinfo {volume} {484}},\
  \bibinfo {pages} {78} (\bibinfo {year} {2012})}\BibitemShut {NoStop}%
\bibitem [{\citenamefont {Kolobov}\ \emph {et~al.}(1993)\citenamefont
  {Kolobov}, \citenamefont {Davidovich}, \citenamefont {Giacobino},\ and\
  \citenamefont {Fabre}}]{PhysRevA.47.1431}%
  \BibitemOpen
  \bibfield  {author} {\bibinfo {author} {\bibfnamefont {M.~I.}\ \bibnamefont
  {Kolobov}}, \bibinfo {author} {\bibfnamefont {L.}~\bibnamefont {Davidovich}},
  \bibinfo {author} {\bibfnamefont {E.}~\bibnamefont {Giacobino}}, \ and\
  \bibinfo {author} {\bibfnamefont {C.}~\bibnamefont {Fabre}},\ }\href
  {\doibase 10.1103/PhysRevA.47.1431} {\bibfield  {journal} {\bibinfo
  {journal} {Phys. Rev. A}\ }\textbf {\bibinfo {volume} {47}},\ \bibinfo
  {pages} {1431} (\bibinfo {year} {1993})}\BibitemShut {NoStop}%
\bibitem [{\citenamefont {Kuppens}\ \emph {et~al.}(1994)\citenamefont
  {Kuppens}, \citenamefont {van Exter},\ and\ \citenamefont
  {Woerdman}}]{PhysRevLett.72.3815}%
  \BibitemOpen
  \bibfield  {author} {\bibinfo {author} {\bibfnamefont {S.~J.~M.}\
  \bibnamefont {Kuppens}}, \bibinfo {author} {\bibfnamefont {M.~P.}\
  \bibnamefont {van Exter}}, \ and\ \bibinfo {author} {\bibfnamefont {J.~P.}\
  \bibnamefont {Woerdman}},\ }\href {\doibase 10.1103/PhysRevLett.72.3815}
  {\bibfield  {journal} {\bibinfo  {journal} {Phys. Rev. Lett.}\ }\textbf
  {\bibinfo {volume} {72}},\ \bibinfo {pages} {3815} (\bibinfo {year}
  {1994})}\BibitemShut {NoStop}%
\bibitem [{\citenamefont {Meiser}\ and\ \citenamefont {Holland}(2010)}]{MEH10}%
  \BibitemOpen
  \bibfield  {author} {\bibinfo {author} {\bibfnamefont {D.}~\bibnamefont
  {Meiser}}\ and\ \bibinfo {author} {\bibfnamefont {M.~J.}\ \bibnamefont
  {Holland}},\ }\href {\doibase 10.1103/PhysRevA.81.033847} {\bibfield
  {journal} {\bibinfo  {journal} {Phys. Rev. A}\ }\textbf {\bibinfo {volume}
  {81}},\ \bibinfo {pages} {033847} (\bibinfo {year} {2010})}\BibitemShut
  {NoStop}%
\end{thebibliography}
